\documentclass[prl,final,twocolumn,showpacs,showkeys]{revtex4}
\usepackage{amsmath}
\usepackage{graphicx}
\usepackage{amsfonts}
\usepackage{amssymb}
\usepackage{subfigure}

\begin{document}



\title{Parallel State Transfer and Efficient Quantum Routing on Quantum Networks}

\author{Christopher Chudzicki}
\author{Frederick W. Strauch} \email[Electronic address: ]{Frederick.W.Strauch@williams.edu}
\affiliation{Williams College, Williamstown, MA 01267, USA}
\date{\today}

\begin{abstract}
We study the routing of quantum information in parallel on multi-dimensional networks of tunable qubits and oscillators.  These theoretical models are inspired by recent experiments in superconducting circuits using Josephson junctions and resonators.  We show that perfect parallel state transfer is possible for certain networks of harmonic oscillator modes.  We further extend this to the distribution of entanglement between every pair of nodes in the network, finding that the routing efficiency of hypercube networks is both optimal and robust in the presence of dissipation and finite bandwidth.

\end{abstract}
\maketitle

The rapid development of coherent superconducting quantum bits (qubits) \cite{Clarke2008} demands continued theoretical analysis of quantum networks.  Current experiments \cite{DiCarlo2010} are in fact realizing multi-dimensional coupling topologies, and advanced coupling networks are on the horizon \cite{Strauch2008,Helmer09}.  While these can provide the hardware to run quantum algorithms, they can also be seen as the fabric of a quantum network to route quantum information between arbitrary nodes in parallel.  This network can be used for the transfer of entanglement, which can be locally purified \cite{Bennett96} for teleportation, error correction \cite{Knill2005} or other tasks in quantum information processing.  

Efficient routing of entanglement between nodes can be be accomplished provided two conditions are met.  First, it must be possible to program the network to transfer quantum states between arbitrary nodes.  This is possible if the qubit frequencies or inter-qubit couplings can be dynamically controlled.  Second, the network should exhibit high fidelity state transfer between nodes in parallel, as a \emph{multi-user} network.  This is in contrast to the various methods studied for state transfer in a \emph{single-user} network; see \cite{Bose2007} for a review.  Here we consider two such networks that meet these conditions---the hypercube and the complete graph, illustrated in Fig. 1.  

To exploit the full parallelism of these networks, we introduce a new approach, using oscillator networks to route multiple excitations between nodes.  Previous theoretical work on quantum routing \cite{Wojcik07} used single state transfer of qubits, while the study of entanglement dynamics in oscillator networks \cite{Plenio04} has commonly focused on single-excitation or continuous-variable entanglement (see \cite{Wu09} for a notable exception).  

In this Letter we theoretically analyze parallel state transfer and quantum routing in these networks.  We begin by providing the first calculations for the fidelity of parallel state transfer, showing that multiple nodes can faithfully send entanglement through the network \emph{at the same time}.   Using these results, we consider entanglement distribution schemes to evaluate the routing efficiency, and provide analytical results for the complete and hypercube networks in the presence of cross-talk and dissipation.  This lays the foundation for analyzing many future quantum networks, and can guide future experiments with superconducting or other qubit-oscillator systems.
   
\begin{figure}[t]
\begin{center}
\includegraphics[width=3.4in]{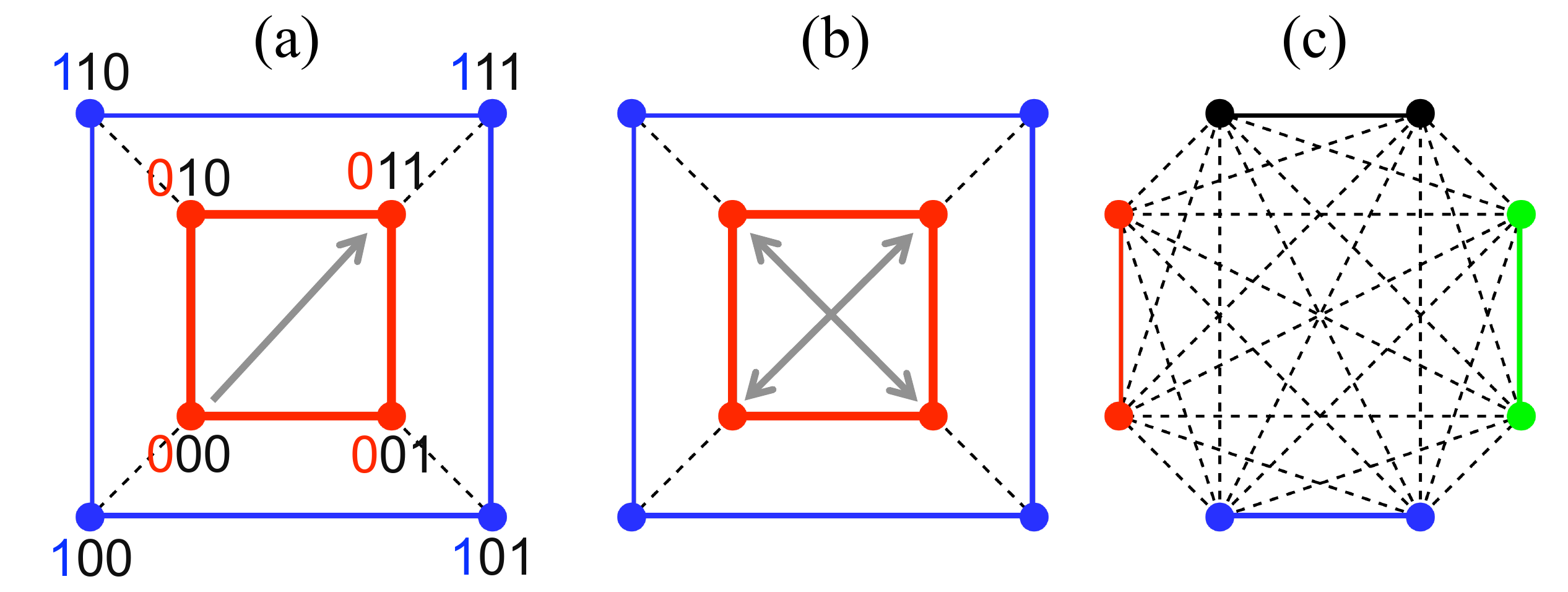}
\end{center}
\caption{Parallel state transfer on programmable quantum networks.  Each node is an oscillator with a tunable frequency.  Each line (solid or dashed) indicates a coupling between oscillators.  Solid lines indicate couplings between oscillators with the same frequency; dashed lines indicate couplings between oscillators with different frequencies.  High fidelity state transfer occurs for large detuning.  (a) Hypercube network with $d=3$, programmed into two subcubes (red and blue squares).  Each node is labeled by a bit-string of length $d=3$, here with the first $m=1$ bits indicating the subcube.  In the qubit-compatible scheme (QC), one entangled pair is sent on each subcube, as indicated by the arrow for the inner (red) square.  (b)  In the massively parallel scheme (MP) scheme, multiple entangled pairs are sent between every node of each subcube, as indicated by the arrows for the inner (red) square.  (b) Completely connected network with $N=8$, programmed into $N/2 = 4$ two-site networks. }
\label{fig1}
\end{figure}

{\it{General framework}}.  A quantum network is described as a graph $\mathcal{G}=(V,E)$ of vertices $V$ connected by edges $E$, with a Hamiltonian $\mathcal{H}$ given by
\begin{equation}
\mathcal{H}= \hbar \sum_{v \in V} \Omega_{vv} a_v^{\dagger} a_v + \hbar \sum_{\{u,v\} \in E} \Omega_{uv}\left( a_u^{\dagger} a_v + a_v^{\dagger} a_u \right).
\end{equation}
where $a_{v}^{\dagger}$ and $a_v$ are the creation and annihilation operators for the oscillator at vertex $v$.  We confine our attention to the case when the couplings between nodes are all equal to $\Omega_0$, although we do require that frequencies of each node $\Omega_{vv}$ be programmable from one use of the network to the next.  Such networks can be realized by tunable superconducting qubits coupled by capacitors or through resonators, as realized in recent experiments \cite{DiCarlo2010}.  While tunable couplings have recently been demonstrated \cite{Allman2010}, their use for quantum routing requires additional study.  Quantum oscillator networks have not yet been realized, but great progress has been made towards qubit-oscillator systems \cite{Sillanpaa2007} and tunable oscillators \cite{Osborn2007}.

To analyze the transfer of entanglement between nodes of a quantum network, we consider the following modification of the state transfer protocol \cite{Bose03,Bose2007}.  At each node of this network there is a register of auxiliary qubits capable of generating pairs of qubits in the Bell state $|\Phi^+\rangle=(|00\rangle+|11\rangle)/\sqrt{2}$.  Some subset of the nodes $\{s_1 \dots s_M\}$ act as senders, and some subset $\{r_1 \dots r_M \}$ act as receivers.  Each sender prepares his auxiliary qubits in a Bell state, and then swaps one of these qubits onto the network (as done in recent experiments \cite{Sillanpaa2007}).  The goal of sender $s_j$ is to faithfully transfer half of her entangled pair to receiver $r_j$.  

More formally, we consider the time-evolution of the initial state
\begin{equation}
|\Psi(t=0)\rangle = 2^{-M/2} \prod_{j=1}^M \left(1 + a_{s_j}^{\dagger} b_{s_j}^{\dagger}\right) |\mbox{vac}\rangle,
\end{equation}
where $b_v^{\dagger}$ is the raising operator for the auxiliary qubit at vertex $v$ and the vacuum state $|\mbox{vac}\rangle$ is the state where all oscillators and qubits are in the ground state.  This initial state is a product of Bell states for each sender's qubit-oscillator pair.  We evolve this state for some time $T$, $|\Psi(T)\rangle = U(T) |\Psi(0)\rangle$ with $U(T) = e^{-i \mathcal{H} T/\hbar}$:
\begin{equation}
|\Psi(T)\rangle = 2^{-M/2} \prod_{j=1}^M \left(1 + U(T) a_{s_j}^{\dagger} U^{\dagger}(T) b_{s_j}^{\dagger}\right) |\mbox{vac}\rangle.
\end{equation}
Our target state has the entanglement transferred to the receiver's node:
\begin{equation}
|\Psi_{\mbox{\scriptsize{target}}}\rangle = 2^{-M/2} \prod_{j=1}^M \left(1 + e^{i \phi_j} a_{r_j}^{\dagger} b_{s_j}^{\dagger} \right)  |\mbox{vac}\rangle,
\end{equation}
where $e^{i \phi_j}$ is a known phase that can be corrected by local unitary operations on the pair $(s_j,r_j)$.  To characterize the state transfer, we use the fidelity of the final density matrix $\rho_{j}$ shared between receiver $r_j$ and the sender $s_j$ (after the appropriate local correction) with our intended perfect Bell state:
\begin{equation}
F_{s_j \to r_j} = \langle \Phi^+ | \rho_{j} |\Phi^+ \rangle.
\end{equation}

We calculate a lower bound on these fidelities by analyzing the time-evolved state and taking its overlap with an appropriately chosen product state $|\Psi_0\rangle = |\Phi^+\rangle_j \otimes |\psi\rangle$.  This calculation is significantly simplified by the linearity of the time-evolution of the oscillator modes, for which
\begin{equation}
U(t) a_j U^{\dagger}(t) = \sum_{k} \mathbb{K}_{jk}(t) a_k,
\end{equation}
where we have defined the unitary mode evolution matrix $\mathbb{K}(t) = \exp(-i \Omega t)$ and where the $N \times N$ coupling matrix $\Omega$ has elements $\Omega_{uv}$.  Note that perfect state transfer would occur for $U (T) a_{s_j}^{\dagger} U^{\dagger}(T) = e^{i \phi_j} a_{r_j}^{\dagger}$.

{\it{Parallel state transfer on the hypercube}}.  Christandl {\it et al.} showed that one can perform perfect state transfer from corner-to-corner of a $d$-dimensional hypercube in constant time $T = \pi/(2 \Omega_0)$ \cite{Christandl04}.  Here we analyze how this result can be extended to the transfer of quantum states in parallel, by splitting the cube into subcubes.  Specifically, by tuning the frequencies of each node, the $d$-dimensional hypercube can be broken up into $2^m$ subcubes each of dimension $d-m$, as shown in Fig. \ref{fig1}(a) for $d=3$ and $m=1$. These subcubes can be made to act as good channels between \emph{their} antipodal nodes by separating the oscillator frequencies for each channel from adjacent channels by an amount $\Delta \omega$.  For fixed couplings, there is still the potential for cross-talk between channels, which we now analyze.

Perfect state transfer on the hypercube is most conveniently described using a binary labeling scheme \cite{Strauch2008}, in which each node is labeled by a bit-string of length $d$.  For parallel state transfer, this string separates into $m$ bits that specify the various subcube channels, and $(d-m)$ bits for the position on each subcube, as illustrated in Fig.\ref{fig1}(a).  Using such a labeling, we choose the detunings so that the coupling matrix takes the form
\begin{equation}
\Omega = \sum_{j=1}^{m}  \left( \Omega_0 X^{(j)} + \frac{1}{2} \Delta \omega Z^{(j)} \right) + \Omega_0 \sum_{j=m+1}^d X^{(j)},
\label{hypcoupling}
\end{equation}
where $X^{(j)}$ and $Z^{(j)}$ are Pauli matrices for the $2^d$ nodes of the graph.  This coupling matrix has the nice property that all oscillators on the same subcube have the same frequency, and are detuned from adjacent subcubes by $\Delta \omega$.  Crucially, we have written Eq. (\ref{hypcoupling}) as a sum of $d$ commuting matrices.  For $\Delta \omega \gg \Omega_0$, the first $m$ terms of $\Omega$ keep the subcubes separate, while the remaining $(d-m)$ terms serve to transfer excitations between corners on each subcube.

For this choice of $\Omega$, the mode evolution matrix $\mathbb{K}(t) = \exp(-i \Omega t)$ can be evaluated exactly.  Then, by choosing the $M = 2^m$ senders and receivers to transfer in the same direction along each of their subcubes (e.g. from $00 \cdots 0$ to $11 \cdots 1$), one can show \cite{ChudzickiThesis} that the fidelity satisfies
\begin{equation}
\label{ES:fid}
F_{s_j \to r_j} \ge 1 - \frac{3}{2} m \ \eta^2 \sin^2 \xi_T + \mathcal{O}(\eta^3),
\end{equation}
for all $j$, where $\eta = 2 \Omega_0/\Delta \omega$ and $\xi_T = \sqrt{1+\eta^{-2}}$.  Note that for large detuning ($\eta \ll 1$) each sender transfers entanglement to her corresponding receiver with high fidelity.  In addition, the fidelity exhibits resonances due to the dependence on $\sin^2 \xi_T$; by picking $\eta$ just right, one can make $\sin^2 \xi_T =0$.  Remarkably, the exact calculation shows that for these detunings the fidelity is equal to one to all orders in $\eta$, realizing perfect parallel state transfer.

\begin{figure}[t!p]
  \begin{center}
		\includegraphics[width=3in]{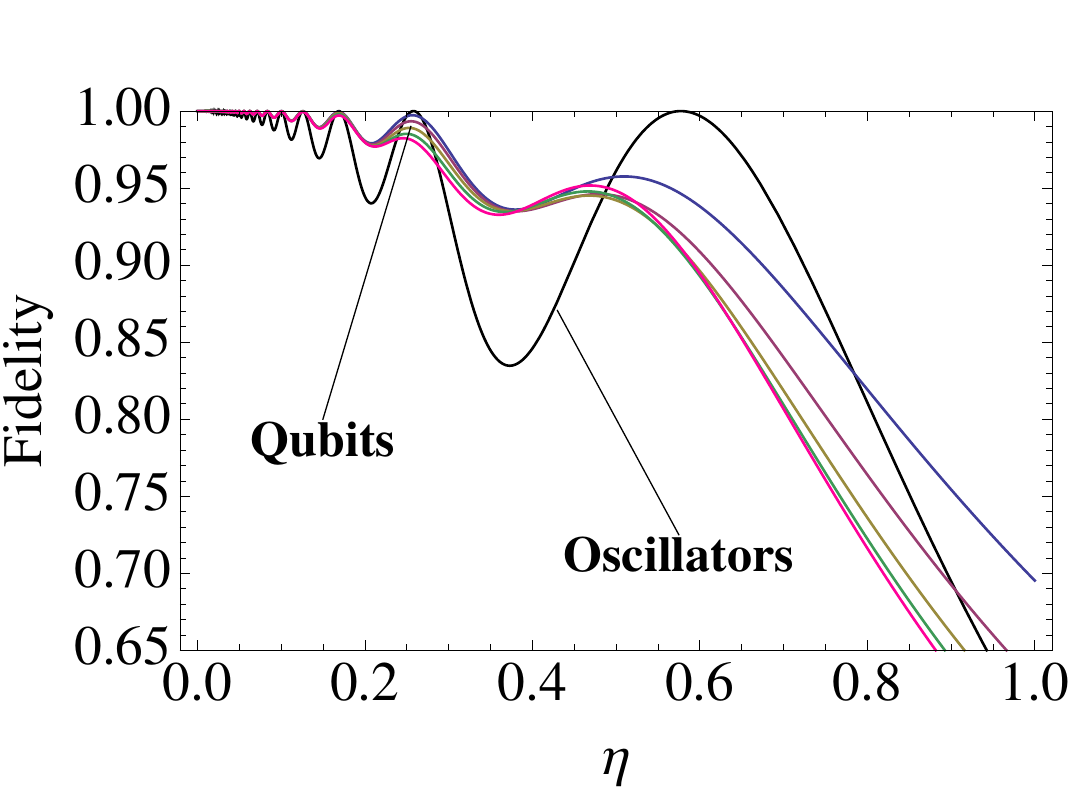}
  \end{center}
\caption{ The fidelity of entanglement transfer on the hypercube as a function of the detuning parameter $\eta = 2 \Omega_0/\Delta \omega$.  The qubit curves are, from top to bottom (for small $\eta$), numerical simulations for dimension $d=2 \to 6$.  Each network is split into $M=2$ subcube channels with one sender and receiver per channel, with entanglement being sent in the same direction on each channel.  Also shown also is the lower bound of Eq. (\ref{ES:fid}) for the oscillator network with $m=1$.}
  \label{fig2}
\end{figure}

Equation \eqref{ES:fid} was derived for entanglement transfer on oscillator networks.  However, as long as only one sender-receiver pair uses each channel at a time, numerical calculations, shown in Fig. \ref{fig2}, indicate that qubit networks behave similarly.  For this reason we call the parallel state transfer protocol discussed so far the ``qubit-compatible'' (QC) protocol.  There are some notable differences between qubits and oscillators, namely qubits do better on average, but do not exhibit perfect state transfer.

Besides the ability to perform truly perfect parallel state transfer, oscillator networks have another feature that qubit networks lack: the capacity for massively-parallel (MP) entanglement transfer. Because multiple excitations on an oscillator network pass through each other without interacting, \emph{every} node on an oscillator hypercube network can act as both sender and receiver at once, with each transferred pair having a fidelity given by Eq. \eqref{ES:fid}.  That is, instead of sending $M=2^m$ states in parallel (as described above), one can send $2^d$ states all at once, by transferring $2^{d-m}$ states on each of the $2^m$ subcubes.  This is illustrated by the arrows in Fig.~\ref{fig1}(b) for $d=3$ and $m=1$, with $4$ states transferred on each square.  In this scheme, all of the network is used all of the time.  The fidelity of each transfer is precisely the same as in the QC scheme.   As we will soon show, this allows for optimal routing efficiency on the hypercube.

{\it{Parallel state transfer on the complete graph}}.  The complete graph is intrinsically qubit-compatible and massively parallel, as all pairs are simultaneously coupled.  Such a network can be realized by coupling many superconducting qubits to a common metallic island or a high-frequency superconducting resonator \cite{DiCarlo2010}.  One can address different pairs of qubits by tuning them into resonance, and separated from other pairs by $\Delta \omega$.  Using perturbation theory on such a configuration, one finds that 
\begin{equation}
F_{\mbox{\scriptsize{complete}}} \ge1 - \frac{\pi^2}{2} \eta^2 + \mathcal{O}(\eta^3),
\label{fcomplete}
\end{equation}
where again $\eta = 2 \Omega_0/\Delta \omega$.   In contrast with the hypercube, here there is no dependence on the number of nodes, and the optimal transfer time is in fact different for different pairs.  The latter complication is ignored in the following.  However, the dependence on the bandwidth through $\eta$ plays an important role in the routing efficiency.

{\it{Entanglement Distribution and Routing Efficiency}}.  Using parallel state transfer, these networks can be programmed for the efficient transfer of quantum information.  To quantify this efficiency, we consider the distribution of entanglement between every node.  That is, we consider protocols that route entanglement between every pair of nodes on the network.  In addition, we consider the realistic condition that all oscillator frequencies are within a finite bandwidth.  We now calculate the routing efficiency for our three protocols.

To quantify the routing efficiency, we use the rate of entanglement distribution.  We define the distribution time $T_D$ as the time it takes for all nodes to share an approximate Bell pair, and the distribution rate as the total number of entangled pairs shared, weighted by their fidelities, divided by the distribution time:
\begin{equation}
\mathcal{R} = \frac{1}{T_D} \sum_{\textrm{pairs $\{u,v\}$}} F_{u \rightarrow v}.
\label{rate}
\end{equation}
This rate distinguishes one-dimensional schemes \cite{Bose03}, that take a long time to transfer a small amount of entanglement, from multi-dimensional schemes that take a short time with perfect state transfer \cite{Christandl04}.  Note that all serial schemes with transfer time $T$ have $\mathcal{R} < 1/T$.  By using parallel state transfer, as described above, this rate can be made even greater.  

For the complete graph, we split the network into $N/2$ pairs.  Each pair is detuned by $\Delta \omega$ from adjacent pairs, for a total bandwidth of $\omega_{\mbox{\scriptsize{max}}} - \omega_{\mbox{\scriptsize{min}}} = N \Delta \omega/2$.  The distribution time $T_D$ is simply the swap time $T = \pi/2\Omega_0$ times the number of steps required to couple each pair of nodes, or $T_D = T (N-1)$.  The total amount of entanglement shared is $N(N-1)$, or two for each pair.  Using the fidelity of Eq. (\ref{fcomplete}), we find that
\begin{equation}
\mathcal{R}_{\mbox{\scriptsize{complete}}} \approx \frac{N}{T} \left(1 - \frac{\pi^2}{2}  \frac{\Omega_0^2}{(\omega_{\mbox{\scriptsize{max}}} - \omega_{\mbox{\scriptsize{min}}} )^2} N^2 \right).
\end{equation}

For the hypercube, we split the network into $2^m$ subcubes.  Here only adjacent subcubes need be detuned, for a total bandwidth of $\omega_{\mbox{\scriptsize{max}}} - \omega_{\mbox{\scriptsize{min}}} = m \Delta \omega$, as indicated in Eq. (\ref{hypcoupling}).  To calculate the distribution time, we consider our two schemes separately.  For the qubit-compatible scheme, in which one excitation is on each subcube channel, the split network must be used for every set of antipodal pairs, or $2^{d-m-1}$ times.  Given that there are $\binom{d}{m}$ such splittings, and that each transfer takes time $T$, the total distribution time is 
\begin{equation}
T_{D}^{\mbox{\scriptsize{(QC)}}} = T \sum_{m=0}^{d-1} \binom{d}{m} 2^{d-m-1} = \frac{1}{2} (3^d-1) T.
\end{equation}
Using Eq. (\ref{ES:fid}) with $\sin^2 \xi_T = 1$, the sum in Eq. (\ref{rate}) can be performed analytically, from which we find, for large $d$,
\begin{equation}
 \mathcal{R}^{\mbox{\scriptsize{(QC)}}} \approx  \frac{1}{T} N^{0.415} \left( 1 - \frac{3}{4} \frac{\Omega_0^2}{(\omega_{\textrm{max}}-\omega_{\textrm{min}})^2}  d^2 (d+3) \right),
\end{equation}
where $N=2^d$ and the prefactor is $(4/3)^d = N^{\log_2(4/3)}$.    

For the massively parallel scheme, every antipodal pair transfers simultaneously in time $T$, giving
\begin{equation}
T_{D}^{\mbox{\scriptsize{(MP)}}} = T \sum_{m=0}^{d-1} \binom{d}{m} = (2^d-1) T,
\end{equation}
with a rate 
\begin{equation}
 \mathcal{R}^{\mbox{\scriptsize{(MP)}}} \approx  \frac{1}{T} N \left( 1 - \frac{3}{4}  \frac{\Omega_0^2}{(\omega_{\textrm{max}}-\omega_{\textrm{min}})^2}  d^2 (d+3) \right).
\end{equation}
These rates are conservative estimates; higher rates are possible by using the resonances seen in Fig.~\ref{fig2}.

\begin{figure}[tb]
  \begin{center}
 \includegraphics[width=3in]{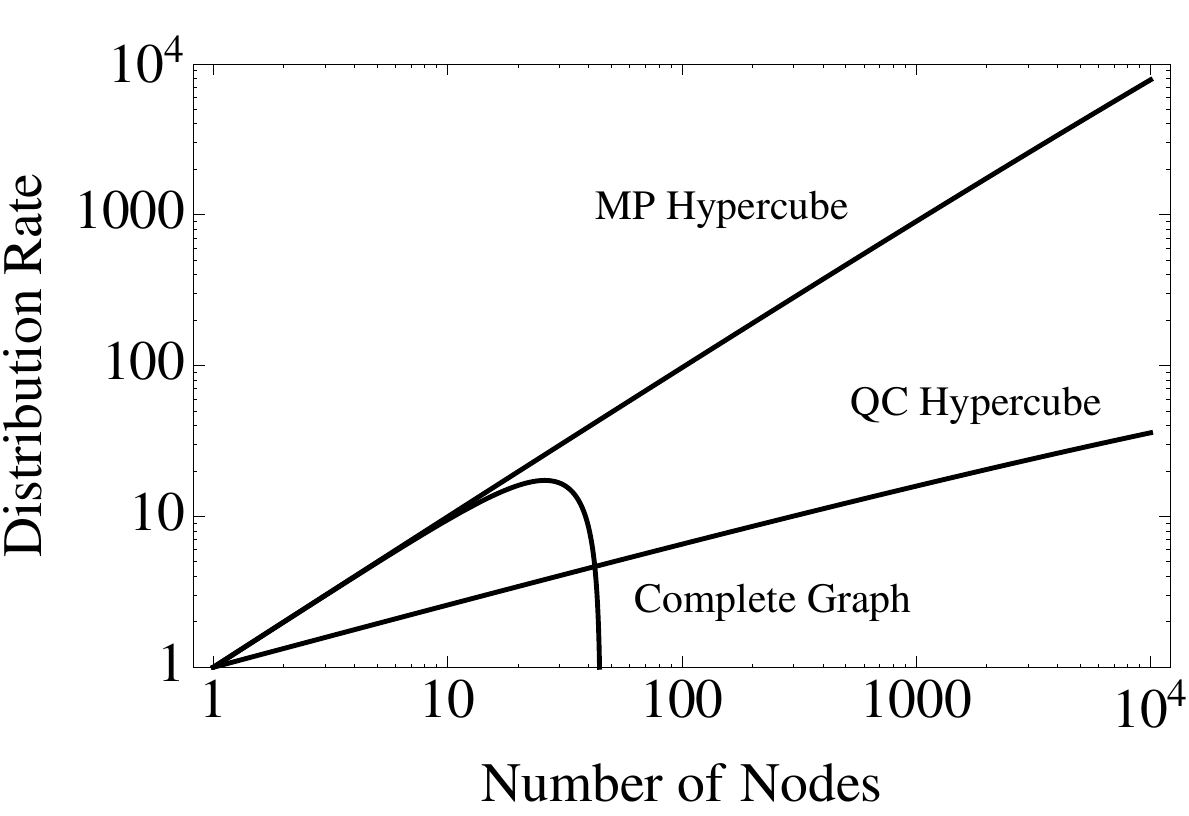}
  \end{center}
\caption{Entanglement distribution rate $\mathcal{R}$ (in units of $1/T$) as function of the number of nodes $N$ in a quantum network.  Three distribution schemes are shown: the massively parallel (MP) and qubit-compatible schemes on the hypercube of dimension $d$ (each with $N=2^d$), and the complete graph of size $N$.  Each network was chosen to to have a coupling of  $\Omega_0/2\pi = 20 \mbox{MHz}$ with a bandwidth $(\omega_{\textrm{max}}-\omega_{\textrm{min}})/2\pi = 2 \mbox{GHz}$. }
  \label{fig:ESplot}
\end{figure}

These three distribution rates are plotted as a function of the number of nodes in Fig. \ref{fig:ESplot}, where we have fixed the bandwidth appropriate to recent superconducting qubit experiments \cite{DiCarlo2010}.  For the hypercube schemes, the massively parallel protocol is more than quadratically better than the qubit-compatible scheme.  Entanglement transfer on the complete graph quickly fails due to significant cross-talk for $N \approx 20$.  One might expect that this is due to the large number of couplings, but it is actually due to the finite bandwidth of the network.  It is clear that studying extended coupling schemes such as the cavity grid \cite{Helmer09} is an important task.

{\it{Decoherence and Disorder for the Hypercube}}.  Experimental issues related to hypercube state transfer, including decoherence and disorder, have been analyzed previously \cite{Strauch2008}.  These results can be applied directly to the QC scheme.  Decoherence will simply reduce the fidelity (and $\mathcal{R}$) by a factor $\sim e^{-T/T_2}$ for arbitrarily large subcubes \cite{Strauch2008}, where $T_2$ is the total dephasing time.   A full analysis of decoherence for the multiple excitations of the MP scheme will require additional study, but the dominant source of decoherence for superconducting oscillators is dissipation \cite{Wang2009}.  For this decoherence process, one can calculate that the fidelity will be reduced by a factor $\sim e^{-T/T_1}$, where $T_1$ is the dissipation time.

In conclusion, we have analyzed how entanglement can be routed between all nodes of the hypercube and completely connected networks.  This has been accomplished by parallel state transfer and analytical calculations of the entanglement distribution rate.  In the ideal case, oscillators on both the hypercube and complete graphs achieve optimal efficiency.  This efficiency is robust for the hypercube in the presence of finite bandwidth and dissipation.  These results provide further evidence that superconducting resonators are an important element for quantum information processing, and motivate further study of parallelism in quantum networks.

\acknowledgments
We gratefully acknowledge discussions with W. Wootters, L. Bishop, K. Jacobs, D. Schuster, and R. W. Simmonds.  This work was supported by the Research Corporation for Science Advancement, and CC by NSF grant PHY-0902906.

\bibliography{report}

\end{document}